\documentclass[a4paper,12pt]{article}
\begin{document}
\hspace{11cm} 
\vspace {1.5cm}

\begin{center}
\noindent {\bf Some remarks on a paper by G.-S. Yang and H.-C.Kim:\\ Mass splitting of SU(3) baryons within a chiral soliton model}
\end{center}
\vskip 13 pt
\begin{center}
G.Morpurgo
\vskip 3 pt
Istituto Nazionale di Fisica Nucleare (Universita` di Genova-Italy).\footnote{e-mail morpurgo@ge.infn.it}\\
\end{center}
\vskip 20 pt
\noindent {\bf Abstract.}
In a recent paper Yang and Kim discuss the lowest baryon mass formulas using a chiral soliton model. However, contrary to 
their assertion, their result is not equivalent to a generalized Gell Mann-Okubo mass formula derived (G.Morpurgo,1992)
using only general properties of QCD (The General Parameterization method). The reason for the non 
equivalence is that the generalized mass  formula includes all the main terms that break flavor at second order,
whereas the Authors above do not go beyond first order flavor breaking (and should therefore find 
the usual Gell Mann Okubo formula). This is not so, due to some inconsistencies. Below we confine to the part
of the Yang-Kim paper dealing with the mass formula. We will not consider the other items of their paper.
\vskip 15pt

\noindent {\bf 1.Second order versus first order in flavor breaking.}\\
As stated in the Abstract, I will consider only the part of the paper of Yang and Kim \cite{yk} dealing
with the Gell Mann-Okubo mass formula (see the generalized Gell Mann-Okubo mass formula Eqs.(1) below) 
derived \cite{mo} many years ago. In the Eqs.(1,2) below the particle symbols stay for the particle masses.\\
\begin{equation} 
\label{eq:A}
   \frac{1}{2}(p+\Xi^{0})+ T = \frac{1}{4}(3\Lambda
   +2\Sigma^{+}-\Sigma^{0})\hspace{2cm}
\end{equation} 
where $T$ stays for
\begin{equation} \label{eq:second}
    T\equiv\Xi^{*-} - \frac{1}{2}(\Omega +\Sigma^{*-})
   \end{equation}
The fit to the data depends on whether we use the conventional or the pole values for the masses of
the resonances appearing in $T$:
\begin{eqnarray}&& 
l.h.s.= 1132.36\pm 0.7\, MeV, \quad \quad  r.h.s.= 1133.93\pm\, 0.04 MeV \,\quad\quad (conventional)\nonumber\\
&&l.h.s.= 1133.86\pm 1.25\, MeV,\quad \quad  r.h.s.= 1133.93\pm\, 0.04 MeV \quad\quad (pole)
\end{eqnarray} 
\indent In Eq.(1) the value of $T$ is $5.7\pm 0.7\,MeV$. The Eq.(1), 
 written with the signs of the charges indicated in it, includes correctly the effects of the e.m. 
 contributions to the masses.
(See in Ref.[2], the text and the footnote 9). Note from Eq.(3) that the accuracy of the above formulas 
is significantly lower than, but recalls that of the Coleman-Glashow relation \cite{3} for the baryon masses.
(Compare also Ref.\cite{4}).\\
\indent As stated in the Abstract, the Eq.(1) was derived by the general QCD parametrization (GP)\cite{5}
 (for a recent review of the GP compare \cite{rdm}). The GP exploits only very general properties of QCD. 
  Note that Eq.(1) was re-obtained several years later by Durand et al. using a chiral QCD
 procedure (see \cite{6}, \cite{7}); their result, however, did not consider the problems related to 
 the e.m. contributions to the masses.\\
 \indent We now consider the paper of Yang and Kim \cite{yk} and start with their Eq.(30):
 \begin{equation}
  2\Xi^{*-} - \Sigma^{*-}=\Omega^{-}
 \end{equation}
 If this equation (one of the equations derived by Gell Mann \textit{at first order in flavor breaking})
 were rigorously correct, our $T$ (Eq.(2)) would clearly vanish. This would imply  the vanishing of any
 correction to the Gell Mann-Okubo 
 formula. Not only this is obviously incorrect (second order flavor breaking is there!), but, as stated, it
  would contradict the results of QCD.\\
 \indent The reason why Yang and Kim took their Eq.(30) as one of the pillars in their treatment is not clear. 
 Possibly they took the first order flavor breaking seriously and used their Eq.(30) because they did not
 intend to go beyond it. Another possibility (because Yang and Kim refer often to a paper of
 Guadagnini \cite{g}) is that they obtained their Eq.(30) from the three last equations of 
 formula (5.4) of that paper [its title is "Baryons as solitons and mass formulas"]. But in that paper
 Guadagnini apparently limits to first order in flavor breaking, whereas Yang and Kim assert 
 incorrectly (compare the statement above their Eq.(33)).
 that "their formulas are basically the same as those of Morpurgo \cite{mo}".\\
\indent To conclude, we summarize the above remarks stating that the Yang Kim formulas (their Eqs.(32))
 -first order flavor breaking- may be the same as the Guadagnini formula, but have
  nothing to do with the generalized Gell Mann Okubo mass formula derived in \cite{mo} and discussed above.\\
\indent I am very grateful to Prof. G.Dillon for a clarifying conversation.
 

\small \baselineskip 8pt
\begin{thebibliography} {10}
\bibitem{yk} G.-S.Yang and H.-C. Kim, arXiv: 1010.3782v2 [hep-ph]
\bibitem{mo} G.Morpurgo, Phys.Rev.Lett.,{\bf68}, 139 (1992)
\bibitem{3} S.Coleman and S.L.Glashow, Phys.Rev.Lett. {\bf6}, 423 (1961)
\bibitem{4} G.Dillon and G.Morpurgo, Phys.Lett.B {\bf481},239 (2000) and (E) Phys.Lett.B {\bf485},429 (2000)
\bibitem{5} G.Morpurgo, Phys.Rev. D {\bf40}, 2997 (1989); G.Morpurgo, Phys.Rev. D {\bf41}, 2865 (1990). (The second of these refs. 
shows explicitly -in its last Sect.- how, for mesons, the GP includes terms of the Sacharov-Zeldovich type or DGG type. The same 
is true for baryons).
\bibitem{rdm} G.Dillon and G.Morpurgo, La Rivista del Nuovo Cimento, {\bf33}, N.1 p.1-55, (2010)
\bibitem{6} L.Durand ,P.Ha and G.Jaczko, Phys.Rev. D {\bf65}, 034019 (2002) and (E) Phys.Rev.D, {\bf65}, 099904
\bibitem{7}L.Durand and P.Ha, Phys.Rev.D {\bf67} 07317, 2003; see also: G.Dillon and G.Morpurgo, Phys.Rev. D {\bf68}, 014001, 2003; 
\bibitem{g} E.Guadagnini, Nucl.Phys.B {\bf236}, 43 ,(1984).
\end{thebibliography}
\end{document}